\RequirePackage{amsmath}
\documentclass{article}
\usepackage{arxiv}
\usepackage{microtype}
\usepackage{graphicx}
\usepackage{amssymb}

\usepackage[caption=false]{subfig}
\usepackage{multirow}
\usepackage{hyperref}
\usepackage{xcolor}
\usepackage{cite}
\usepackage{tikz}
\usetikzlibrary{positioning}
\usepackage[round-mode=places, round-integer-to-decimal, round-precision=5,
table-format = 1.5,
table-number-alignment=center,
round-integer-to-decimal]{siunitx}

\newcommand{\pre}{{\text{DP}}}
\newcommand{\post}{{\text{RP}}}
\newcommand{\corrupted}{{\text{cor}}}

\def\etal{\emph{et~al}.}

\usepackage{authblk}
\author[1,2]{ Michael Rotman}
\author[2]{Rafi Brada}
\author[2]{Israel Beniaminy}
\author[3]{Sangtae Ahn}
\author[3]{Christopher J. Hardy}

\author[1]{Lior Wolf }

\affil[1]{School of Computer Science, Tel-Aviv University, Tel-Aviv, Israel}
\affil[2]{GE Global Research, Israel}
\affil[3]{GE Global Research, Niskayuna}
\date{}

\begin{document}
\title{A Novel Approach for Correcting Multiple Discrete Rigid In-Plane Motions Artefacts in MRI Scans}
\maketitle          
\begin{abstract}
	Motion artefacts created by patient motion during an MRI scan occur frequently in practice, often rendering the scans clinically unusable and requiring a re-scan. While many methods have been employed to ameliorate the effects of patient motion, these often fall short in practice. In this paper we propose a novel method for removing motion artefacts using a deep neural network with two input branches that discriminates between patient poses using the motion's timing. The first branch receives a subset of the $k$-space data collected during a single patient pose, and the second branch receives the remaining part of the collected $k$-space data.
	The proposed method can be applied to artefacts generated by multiple movements of the patient. Furthermore, it can be used to correct motion for the case where $k$-space has been under-sampled, to shorten the scan time, as is common when using methods such as parallel imaging or compressed sensing. Experimental results on both simulated and real MRI data show the efficacy of our approach.
\end{abstract}

\section{Introduction}
MRI produces images by scanning the spatial frequency domain, $k$-space. MRI scan times tend to be long; a typical brain exam can last between 30 to 60 minutes. Patient motion during the scan may lead to severe imaging artefacts, rendering many scans clinically unusable. While many methods have been employed to ameliorate the effects of patient motion during scanning (surveys are given in \cite{paper:zaitsev,paper:godenschweger}), all have limitations, and motion remains an unsolved problem. One limitation, for instance, is the long processing time for each slice, common to iterative methods~\cite{TAMER}. Recently, Deep Neural Networks (DNN)-based approaches for motion correction have become a focus of research~\cite{paper:moco,paper:retrocorr}. To the best of our knowledge, none of the previous DNN based approaches has utilized motion-timing information, and none were applied to under-sampled $k$-space data.

In this paper we propose a novel DNN-based approach for correcting rigid in-plane patient motion during an MR scan. This is to be distinguished from methods correcting for non-rigid organ motion, e.g. that caused by respiration, peristalsis, or cardiac motion~\cite{paper:cardiac}. Our key contribution is the use of the timing information of the motion which occurred during the scan, without requiring the details of the motion. Motion timing information can be derived for example using optical tracking, navigator methods~\cite{paper:zaitsev,paper:godenschweger} or from the raw multi-coil data~\cite{paper:motiondetection} as was done in this work. The proposed method extracts a self-consistent part of the $k$-space data using the motion timing and the scan order information, to be used for creating a data consistency layer. Both the consistent part and the remaining part of $k$-space are fed into a DNN that computes the corrected image.
The method corrects for artefacts generated by multiple movements of the patient. Furthermore, it can be used to reconstruct and correct for motion artefacts when using under-sampled $k$-space data, a scan mode often used to shorten the scan time.

\section{Background}

\begin{figure}[t]%
	\centering%
	\subfloat[$\mathcal{FS}_{256}$]{%
		\centering%
		\includegraphics[height=0.3\linewidth]{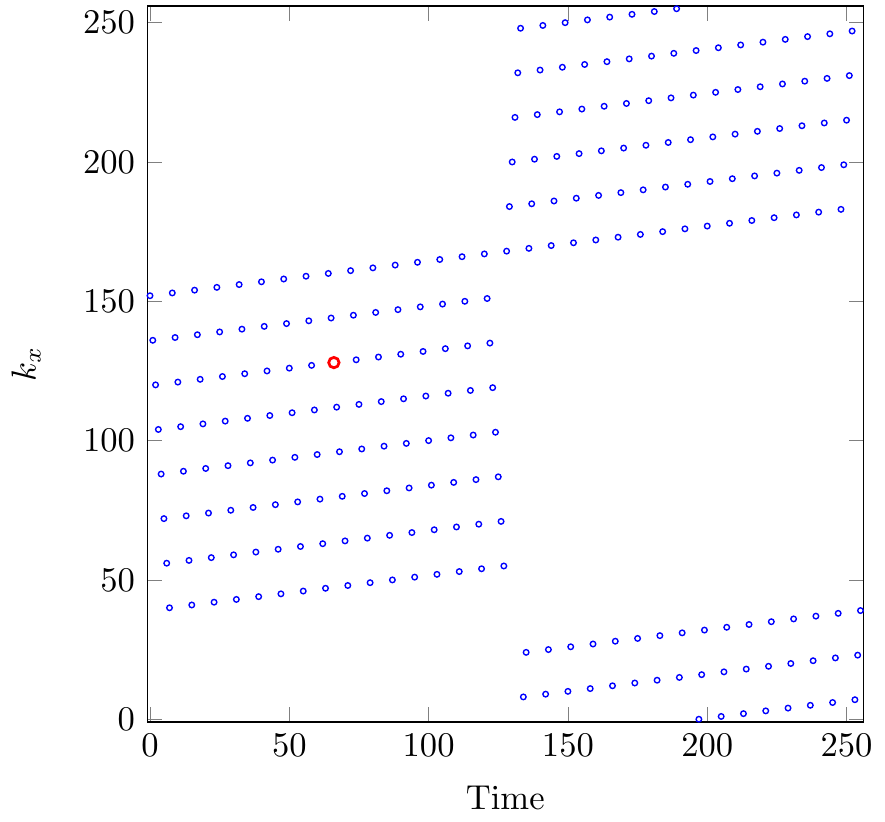}%
		\label{fig:completekscan}%
	}%
	\hfill
	\subfloat[$\mathcal{FS}_{260}$]{%
		\centering%
		\includegraphics[height=0.3\linewidth]{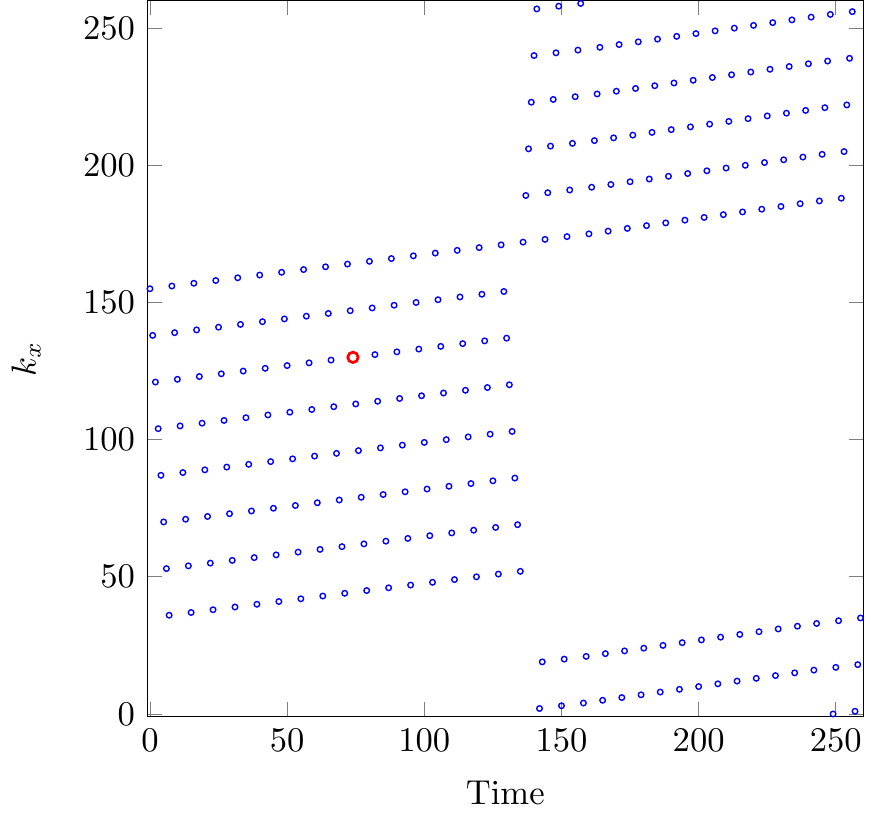}%
		\label{fig:ex51kscan}%
	}%
	\hfill
	\subfloat[${\mathcal{US}}_{260}$]{%
		\centering%
		\includegraphics[height=0.3\linewidth]{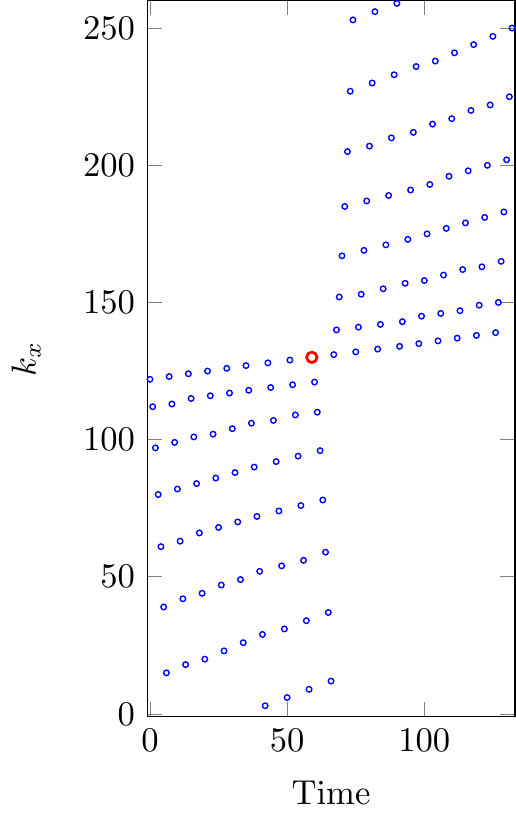}%
		\label{fig:sparsekscan}%
	}
	\caption{The three scan orders (a-c). The $x$ coordinate of each point in this figure represents the time in which that phase-encode ($y$ coordinate) is scanned. We have chosen a convention where $k_x$ is the phase-encode direction, and the readout direction $k_y$ can be visualized as going into the paper. The red points mark $t_{\text{center}}$. (a) The FSE scan order for scanning the fully-sampled $k$-space with spatial resolution of $256\times256$. Low frequency components around the zero-frequency ($k_x=128$) are scanned earlier. (b) The FSE scan order for scanning the fully-sampled $k$-space with spatial resolution of $260\times300$. Low frequency components around the zero-frequency ($k_x=130$) are scanned earlier. (c) The FSE scan order for variable density under-sampling. Low frequency components around the zero-frequency ($k_x=128$) are densely sampled and high frequency components ($k_x$ close to both ends $0$ and $259$) are sparsely sampled, sampling a total of $133$ out of the $260$ lines.}
	\label{fig:scanorders}
\end{figure}

\noindent{\bf Scan Order\enskip}
MRI scanners employ a variety of pulse sequences that utilize a train of RF excitation and magnetic-field gradient pulses, and with the aid of RF receiver coils, collect images that are sensitive to different tissue spin relaxation times. Our work is focused on patient motion correction for a Cartesian 2D multi-slice Fast Spin Echo (FSE) protocol. Data are collected during sequential rastering in the Fourier domain through two-dimensional $k$-space, where a series of columns is measured sequentially for each slice.  As a result, different groups of columns of $k$-space are being scanned at different times.
Each scanning protocol may utilize a different scan order, $\mathcal{S} : T \rightarrow \left\{k_x \right\}$ where $T$ is a set containing the possible timings and $\left\{k_x \right\}$ is the set of the scanned $k$-space columns. In this work we used three scan orders. The first two scan orders, $\mathcal{FS}_{256}$ and $\mathcal{FS}_{260}$, are for a fully-sampled and the last, ${\mathcal{US}}_{260}$, is for an under-sampled $k$-space scanning. The fully-sampled scan orders differ in the number of frequencies acquired; one is intended for acquisition of $256$ columns and the other for $260$ columns in $k$-space. All scan orders are shown in Fig.~\ref{fig:scanorders}. For MR images the lower frequencies contain most of the signal energy. We denote by $t_{\text{center}} \in T$ the the timing in which the zeroth frequency component is scanned.

\noindent{\bf Coil Arrays and Sensitivity Maps\enskip} In order to improve the signal-to-noise ratio and resolution, MRI scanners are equipped with multiple coils. In parallel MR imaging, these coils are also used to reduce the acquisition time using partial $k$-space sampling. For each coil, $i$, there is an associated spatial sensitivity map, $s_i\in\mathbb{C}^{2}$. The image, $x\in\mathbb{C}^2$ can be reconstructed, without any motion correction,  by combining the collected $k$-space data from each coil, $k_i \in \mathbb{C}^2$, 
\begin{equation}
x = \sum_i \left({s_i^{*} \odot \mathcal{F}^{-1}\left\{   k_i\right\}}\right)\,,
\label{eq:ktox}
\end{equation}
where $\odot$ stands for the Hadamard product, $\left(\cdot\right)^*$ is the complex conjugate and $\mathcal{F}$ is the Fourier operator. The sensitivity maps, $s_i$, are normalized such that for each pixel, $(j,k)$, $\sum_i \left\vert  s_i\left(j,k\right) \right\vert^2 = 1$.

\noindent{\bf Rigid-Body In-Plane Motion Artefacts\enskip}
Patient motion may produce motion artefacts in the reconstructed image. For simplicity, we assume that the movement is in-plane, that it takes a much shorter time than the scan time, and that the scanned anatomy behaves as a rigid body. For the case of a single motion, the measured $k$-space consists of two parts: the first part is the $k$-space data acquired while the patient was in the initial position, and the second part is the $k$-space data acquired while the patient was in the changed position after the motion took place. The inconsistency between the two parts of $k$-space results in artefacts visible in the image. The more evenly the motion timing (when the movement occurs) divides the center region of $k$-space between the two parts, the more severe will be the observed artefacts in the image.

\section{Method}
\label{section:method}

We propose a new method that utilizes the whole signal of the motion-corrupted $k$-space data, $k_i^{\corrupted}$. The first step of our method is to divide the corrupted $k$-space data, $k_i^{\corrupted}$, into two distinct $k$-space components; the first contains the $k$-space data collected from a single patient position, which we term as the Dominant Pose (DP), $k_i^{\pre}$, and the second one, $k_i^{\post}$, contains the Remaining Poses (RP) $k$-space data, which may include multiple motion states. For this purpose we generate an under-sampling mask, $m$ which equals $1$ for columns scanned in the dominant pose and $0$ elsewhere,
\begin{eqnarray}
	k_i^{\pre} &=& m \odot k_i^{\corrupted} \\
	k_i^{\post} &=& \left(1 -m\right) \odot k_i^{\corrupted} \,.
\end{eqnarray}
The DP is selected to be the motion state (pose) that includes the larger part of the frequencies near the center of $k$-space.

Both $k_i^{\pre}$ and $k_i^{\post}$ are transformed separately into image space using Eq.~\eqref{eq:ktox},
\begin{eqnarray}
x^{\pre}_0 = \sum_i \left({s_i^{*} \odot \mathcal{F}^{-1}\left\{   k_i^{\pre}\right\}}\right) &,&\quad x^{\post}_0 = \sum_i \left({s_i^{*} \odot \mathcal{F}^{-1}\left\{   k_i^{\post}\right\}}\right) \,,
\end{eqnarray}
 and are used as the two channel input (real and imaginary) to the two branches, $x^{\pre}_0$ and $x^{\post}_0$, a main and a secondary branch of a cascaded neural network. The cascaded neural network consists of $10$ identical units, $\mathcal{U}_n$, $n=1\dots10$, adapting Schlemper \etal~\cite{paper:schlemper}. In each unit, a ResNet~\cite{paper:resnet} block $R_n$, is applied to the $x^{\post}_{n-1}$ in order to extract additional information from the remaining poses,
\begin{equation}
x_{n}^{\post} = R\left(x_{n-1}^{\post}\right) \,.
\end{equation}
The transformed RP complex image is concatenated with the DP complex image, $x_{n-1}^{\pre}$, to form a four channel input to network $U_n$, a variant of the U-Net~\cite{paper:unetpaper} architecture, containing $6$ levels of down and up-sampling,
\begin{equation}
\tilde x_{n}^{\pre} =  U_n\left(\left\{x_{n-1}^{\pre}, x_{n}^{\post} \right) \right\}   \,.
\end{equation}

Imposing that the input $k$-space data of the DP part remains the same for $x_{n-1}^{\pre}$ and $x_{n}^{\pre}$ is achieved by the application of a Data Consistency (DC) layer. This layer replaces the $k$-space data at the DP regions using the mask, $m$, together with the original $k$-space data, $k_i^{\pre}$,
\begin{equation}
x_{n}^{\pre} = DC\left(\tilde x_{n}^{\pre},m\right) = \sum_{i}s^*_i\odot\mathcal{F}^{-1}\left\{ k^{\pre}_i+ \left(1-m\right) \odot \mathcal{F}\left\{ \tilde x_{n}^{\pre} \odot s_i \right\} \right\} \,.
\label{eq:DC}
\end{equation}

\begin{figure}[t]
	\centering
	\centering
	\includegraphics[width=1\textwidth]{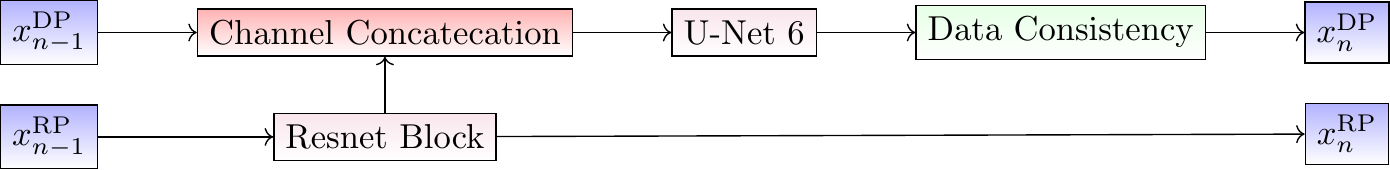}
	\caption{The structure of a single unit, $\mathcal{U}_n$ of the cascaded neural network with two branches. The inputs,		 $x_{n-1}^{\pre}$ and $x_{n-1}^{\post}$, are concatenated after the application of a ResNet Block on $x_{n-1}^{\post}$. The product is then fed to a U-Net, $U_n$, to produce $\tilde x_{n}^{\pre}$ and then altered in the DC layer for the next unit in the cascade.}
	\label{fig:unitstructure}
\end{figure}

The output of the last DC layer, $x^{\pre}_{10}$, is the motion-corrected image. The total computation for each unit, $\mathcal{U}_n$, appears in Fig.~\ref{fig:unitstructure}.

\noindent{\bf Network Optimization\enskip}
\label{section:netopt}
The cascaded neural network is trained end-to-end in a supervised fashion. The inputs, $\mathbb{X}$ are a set of $N$ motion corrupted multi-coil $k$-space data $\left\{k_{i,l}^{\corrupted}  \right\}_{l=1}^N$, and their corresponding motion timings, $\left\{\left\{t_{1l},t_{2l},\dots \right\} \right\}_{l=1}^N$. The targets, $\mathbb{Y}$, are a set of complex images, $\left\{y_{l}^{\pre}  \right\}_{l=1}^N$, reconstructed from the full $k$-space data of the corresponding selected DP.

Optimization is carried out by minimizing the Contrast-weighted SSIM loss~\cite{paper:ssimcontrast},
\begin{equation}
\mathcal{L} = \mathcal{L}_{\text{SSIM-C}}\left(\operatorname{\mathbb{R}e}\left\{x^{\pre}_{10}\right\} ,\operatorname{\mathbb{R}e}\left\{ y^{\pre} \right\} \right) + \mathcal{L}_{\text{SSIM-C}}\left(\operatorname{\mathbb{I}m}\left\{x^{\pre}_{10}\right\} ,\operatorname{\mathbb{I}m}\left\{ y^{\pre} \right\} \right) \,,
\end{equation}
\begin{equation}
\mathcal{L}_{\text{SSIM-C}}\left(x,y\right) = \frac{1}{2} - \frac{1}{2}\left(\frac{2\mu_x \mu_y+ c_1}{\mu_x^2 + \mu_y^2 + c_1}\right)^\alpha \left(\frac{2\sigma_x \sigma_y+ c_2}{\sigma_x^2 + \sigma_y^2 + c_2}\right)^\beta \left(\frac{\sigma_{xy}  +\frac{c_2}{2}}{\sigma_x\sigma_y + \frac{c_2}{2}}\right)^\gamma 
\end{equation}
with the following parameterization,
\begin{equation}
\begin{array}{c}
c_1 = \left(0.01 \max\left\{y\right\}\right)^2\quad
c_2 = \left(0.03 \max\left\{y\right\}\right)^2 \\
\alpha = 0.3 \quad\quad
\beta = 1 \quad\quad
\gamma = 0.3
 \end{array} \,,
\end{equation}
while $\mu_{x,y}$ and $\sigma_{x,y,xy}$ are computed using an $11\times11$ Gaussian kernel with a standard deviation of $1.5$.

\section{Experimental Setup and Results}

\noindent{\bf Data Sources\enskip}
For this study we used two datasets of MRI head scans compiled from various collections. One dataset, which we will refer to as the Diverse dataset, contains $992$ transaxial, sagittal and coronal slices. The second dataset, the Transaxial dataset, contains $7474$ transaxial slices selected from the TCGA-GBA~\cite{dataset:TCGA-GBM} dataset available in the TCIA~\cite{dataset:T1} data collection. The scan resolution in both datasets is $256\times256$ pixels. For the Transaxial dataset we used $8$-coil sensitivity maps computed using the Biot-Savart law\cite{coilsensconstruction}. For the Diverse dataset, we used the measured $8$ coil sensitivity maps obtained from the MRI scanner. For the cases of $\mathcal{FS}_{260}$ and ${\mathcal{US}}_{260}$ scan orders, both the scans and sensitivity maps were padded to a final spatial resolution of $260\times300$, compatible with cases containing real patient motion.

Three subsets were chosen randomly from both datasets:  train (804 samples), validation (86) and test (102) for the Diverse dataset and train (6474), validation (500) and test (500) for the Transaxial one.

\noindent{\bf Data Preparation\enskip}
The column ordering in the scan order, used for the fully-sampled $k$-space cases, $\mathcal{FS}_{256}$ and $\mathcal{FS}_{260}$, is such that the central part of $k$-space data is scanned during the first half of the scan as shown in Fig.~\ref{fig:scanorders}. Accordingly if the time, $t_1$, of the first motion is greater than approximately a quarter of the scan time, the $k$-space data of the DP will be chosen to be the $k$-space data acquired prior to the first motion. As the time, $t_1$, of the first motion increases, the $k$-space data in the DP covers more of the central region of $k$-space.  By contrast, if the first motion occurs earlier than one quarter of the scan time, the amount of $k$-space data in this motion state will be insufficient, causing motion correction to become challenging. In this case the second motion state will be selected as the DP. A similar consideration is used for the under-sampled $k$-space, acquired by ${\mathcal{US}}_{260}$ scan order.

The main aspect of data preparation was the simulation of the patient motions during the scan. For each scan, we draw two first patient motion timings $t_1$, $0 <t_1<t_{\text{center}}$ and $t_{\text{center}} \leq t_1 < \frac{\left\vert T \right\vert}{2}$, since first patient motions occurring after half of $k$-space was scanned lead to weaker motion artefacts and thus are easier to correct for.  For each of these timings we further randomly draw two additional later timings $t_2$ and $t_3$.  A random rigid transformation, $R_i$, is applied for each $t_i$, generating a simulated pose. For the case of $t_1 < t_{\text{center}}$, we introduce an additional constraint requiring a minimal time delay for the second motion, setting  $t_1 + 64  \leq t_2$.

Once the motion parameters are set, the $k$-space data consistent with the simulated poses is calculated. The final corrupted $k$-space data, $k^{\corrupted}_i$ is obtained by merging the respective parts of the different poses' $k$-space data~\cite{paper:lorch2017automated}.

\noindent{\bf Neural Network Structure and Optimization\enskip}
A separate model is trained for each dataset and for each of the scan orders. The Resnet block in each cascade unit contains $3$ convolutional layers with $64$ filters each followed by a LeakyReLU~\cite{paper:leakyrelu} activation function with a slope of $0.2$. We use an adaptation of the U-Net with 6 levels of down-sampling, where the number of filters in each of the convolution layers rises gradually from 64 to 512. In our adaptation there is only one convolutional layer per scale. To evaluate the contribution of merging the RP $k$-space data with the DP $k$-space data, we also defined a Single-Branch model. The ''Single Branch'' model is identical to the Two-Branch model, with the DP $k$-space data replacing the RP $k$-space data as input to the secondary branch, ensuring both models have the same number of parameters. 

Models are trained with a learning rate of $0.0001$ using ADAM~\cite{paper:adam}, for $120$ and $60$ epochs for the Diverse and Transaxial datasets, respectively.

\begin{figure}[t]
	\centering
	\subfloat[]{%
		\centering%
		\includegraphics[height=0.4\linewidth]{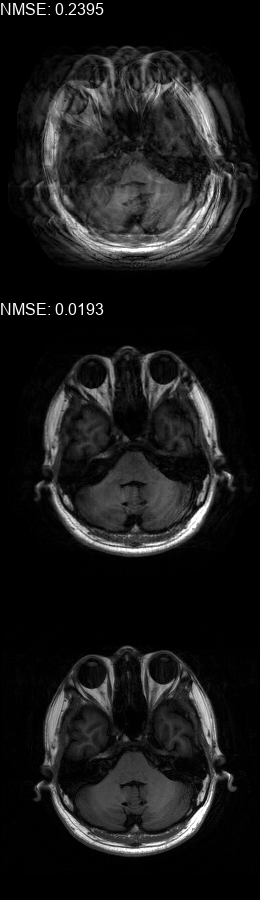}%
		\label{fig:completekscan1}
	}%
	\subfloat[]{%
		\centering
		\includegraphics[height=0.4\linewidth]{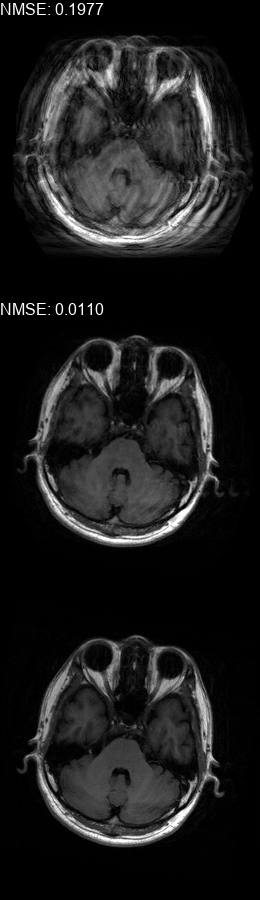}%
		\label{fig:completekscan2}%
	}%
	\subfloat[]{%
		\centering%
		\includegraphics[height=0.4\linewidth]{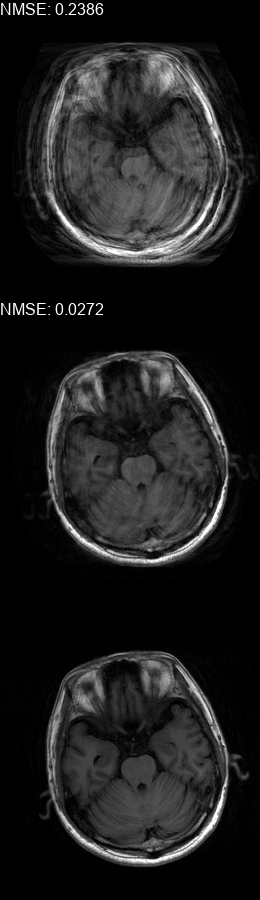}%
		\label{fig:completekscan3}%
	}%
	\subfloat[]{%
		\centering%
		\includegraphics[height=0.4\linewidth]{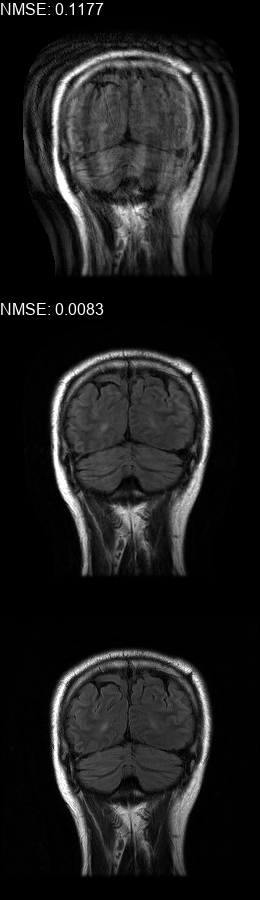}%
		\label{fig:completekscan4}%
	}%
	\subfloat[]{%
		\centering%
		\includegraphics[height=0.4\linewidth]{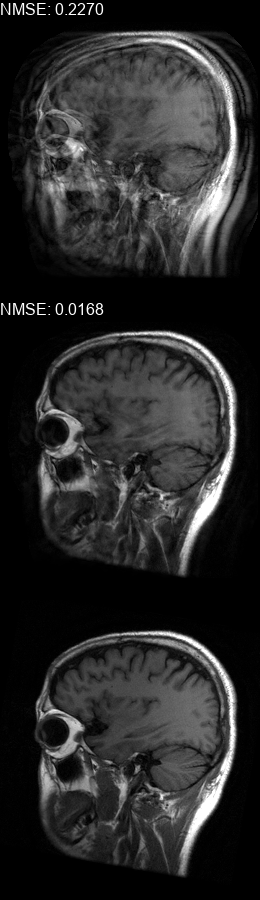}%
		\label{fig:completekscan5}%
	}%
	\subfloat[]{%
		\centering%
		\includegraphics[height=0.4\linewidth]{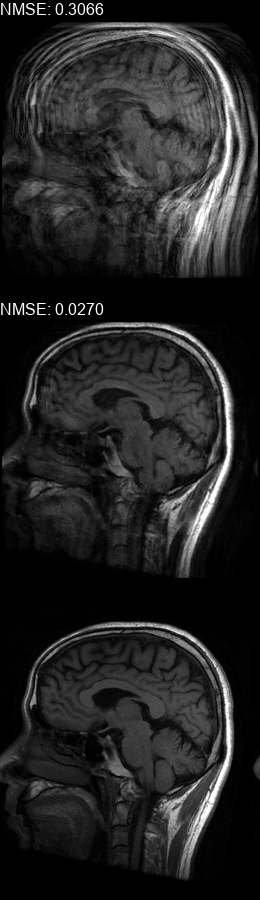}%
		\label{fig:completekscan6}%
	}%
	\caption{%
	Motion corrected images from the Diverse dataset calculated using the  $\mathcal{FS}_{260}$ scan order presented in Fig.~\ref{fig:ex51kscan}. The bottom row shows the motion-free image.  The top row shows the motion-corrupted images. The middle row shows the corrected images calculated using our method. (a) contains a motion at $t_1=87$, (b)  contains two motions at $t_1\!=\!46$, $t_2\!=\! 170$,  (c) contains three motions at $t_1\!=\!72$, $t_2 \!=\! 227$, $t_3\!=\!248$, (d)  contains two motions at $t_1\!=\!112$, $t_2\!=\! 216$, (e) contains a motion at $t_1\!=\!66$, (f) contains three motions at $t_1=65$, $t_2 \!=\! 135$, $t_3\!=\!159$.  }
	\label{fig:completetwobranch}
\end{figure}

\begin{figure}
	\centering
	\subfloat[]{%
		\centering%
		\includegraphics[height=0.4\linewidth]{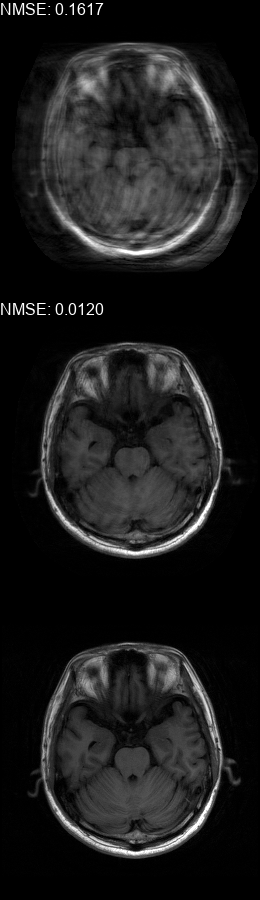}%
		\label{fig:sparsescan1}
	}%
	\subfloat[]{%
		\centering
		\includegraphics[height=0.4\linewidth]{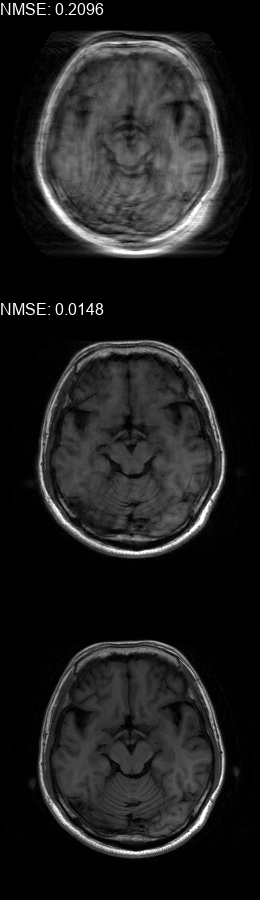}%
		\label{fig:sparsescan2}%
	}%
	\subfloat[]{%
		\centering%
		\includegraphics[height=0.4\linewidth]{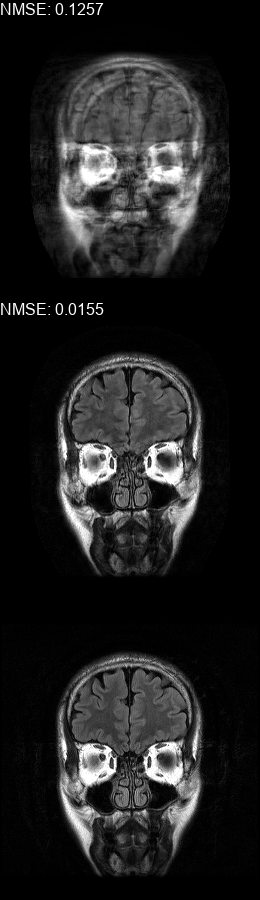}%
		\label{fig:sparsescan3}%
	}%
	\subfloat[]{%
		\centering%
		\includegraphics[height=0.4\linewidth]{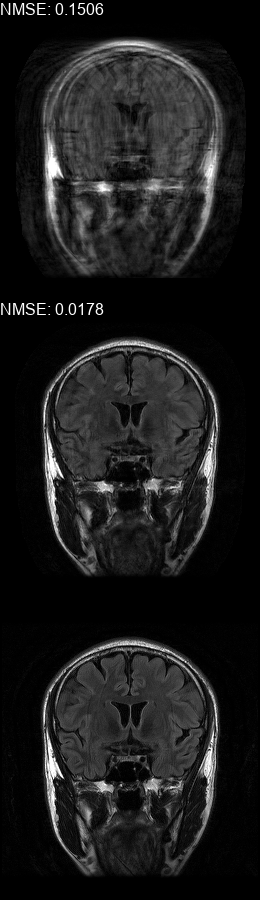}%
		\label{fig:sparsescan4}%
	}%
	\subfloat[]{%
		\centering%
		\includegraphics[height=0.4\linewidth]{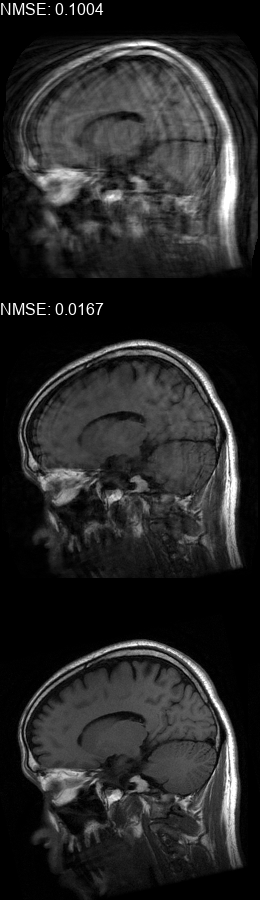}%
		\label{fig:sparsescan5}%
	}%
	\subfloat[]{%
		\centering%
		\includegraphics[height=0.4\linewidth]{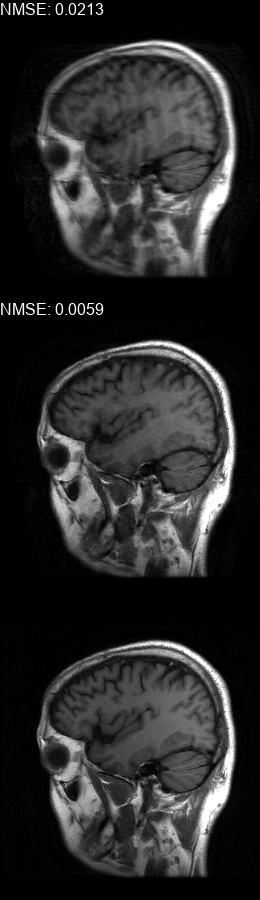}%
		\label{fig:sparsescan6}%
	}%
	\caption{%
		Motion corrected images from the Diverse dataset calculated using the $\mathcal{US}_{260}$ scan order presented in Fig.~\ref{fig:sparsekscan}. The bottom row shows the fully-sampled motion-free images.  The top row shows the under-sampled motion-corrupted images used as input to the correction method, these images show an additional aliasing artefacts in comparison to the ones in the top row of Fig.~\ref{fig:completetwobranch} since they are missing approximately half of the $k$-space data. The middle row shows the corrected images calculated using our method.  (a) contains two motions at $t_1\!=\!79$, $t_2\!=\!132$, (b)  contains a motion at $t_1\!=\!61$,  (c) contains a motion at $t_1\!=\!83$, (d) contains two motions at $t_1\!=\!78$, $t_2\!=\! 123$, (e) contains three motions at $t_1\!=\!13$, $t_2\!=\!97$, $t_3\!=\!127$, (f) contains two motions at $t_1\!=\!55$, $t_2 \!=\!128$. }
	\label{fig:completetwobranchsparse}
\end{figure}

\noindent{\bf Results\enskip}
Experimental results using the two presented datasets, for the fully-sampled and under-sampled $k$-space scan orders, are shown in Table \ref{table:all-results}. The table presents the Normalized Mean Square Error (NMSE) obtained for each of the scan orders. For the Transaxial dataset, we only show the results for the challenging task where the DP contains only up to $70$ columns of $k$-space data. All methods are able to reduce the NMSE of the corrupted scans, however the methods that incorporate the DC layer achieve a much better result.
These results match our subjective visual impression of the motion-corrected images, as may be seen in Fig.~\ref{fig:completetwobranch} for simulated motion in a fully-sampled $k$-space data, and Fig.~\ref{fig:completetwobranchsparse} for an under-sampled $k$-space data. 

\begin{figure}
	\centering%
	\subfloat[]{%
		\centering%
		\includegraphics[height=0.4\linewidth]{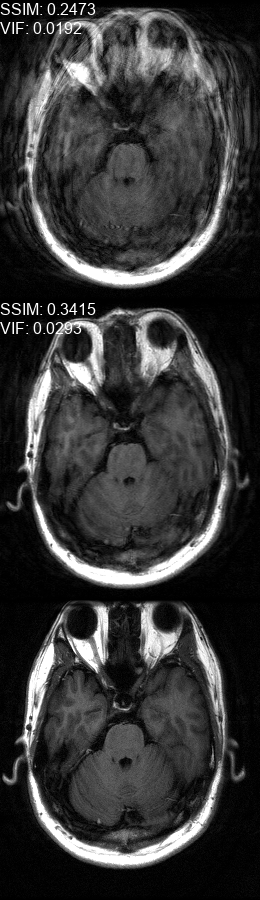}%
		\label{fig:realscan1}
	}%
	\subfloat[]{%
		\centering
		\includegraphics[height=0.4\linewidth]{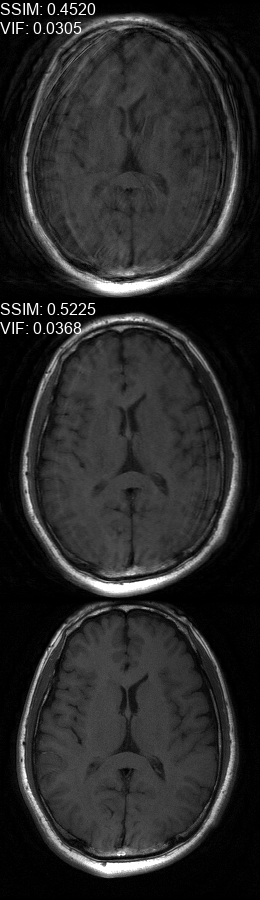}%
		\label{fig:realscan2}%
	}%
	\subfloat[]{%
		\centering%
		\includegraphics[height=0.4\linewidth]{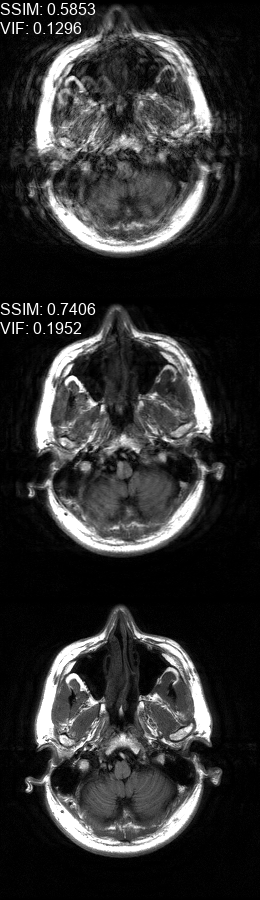}%
		\label{fig:realscan3}%
	}%
	\subfloat[]{%
		\centering%
		\includegraphics[height=0.4\linewidth]{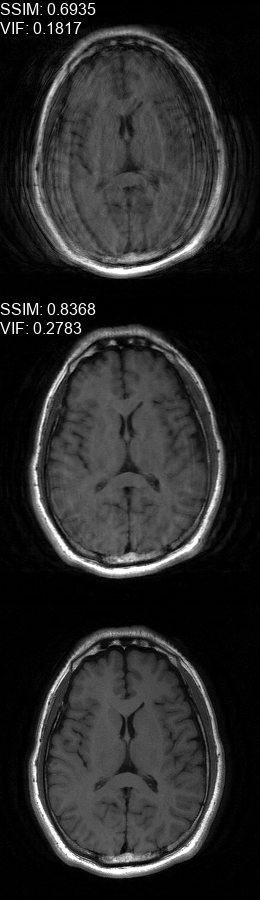}%
		\label{fig:realscan4}%
	}%
	\caption{%
		Examples of real motion corrupted images together with their motion corrected counterparts. The bottom row shows a corresponding slice from a motion-free acquisition of the same subject. These motion-free images were acquired using an additional scan. The images on the top row are motion-corrupted. Images on the middle row were corrected using our method.  (a,b) motion was was detected at $t_1\!=\!96$, (c,d) the motion was detected at $t_1\!=\!112$. All scans were acquired using $\mathcal{FS}_{260}$ presented in Fig.~\ref{fig:ex51kscan}. }
	\label{fig:realcases}
\end{figure}

Fig.~\ref{fig:realcases} contains four scan reconstructions from two patient cases. In both cases, the patient was requested to move at a certain time. In order to provide a ground truth for this scenario, another scan was performed using the same protocol for each patient, where the patient was required to hold still. The motion timing information for this case was obtained using the method described in \cite{paper:motiondetection} as it requires no additional hardware or changes to the scan protocol. We compare the reconstruction error using the SSIM~\cite{ssim} and the Visual Information Fidelity (VIF)~\cite{10.1109/TIP.2005.859378} metrics, since a pixelwise comparison is not viable in this case.

\begin{table}[t]
	\begin{center}
		\caption{NMSE over the different datasets and scan orders.}
		\label{table:all-results}
		\begin{tabular}{|l|p{4em}|p{4em}|p{4em}|p{4em}|p{4em}|p{4em}|}
			\cline{1-7}
			\multicolumn{1}{|l|}{}&  \multicolumn{3}{c|}{Diverse} & \multicolumn{3}{c|}{Transaxial} \\ \cline{2-7}
			\multicolumn{1}{|c|}{Method} &  \multicolumn{1}{c|}{$\mathcal{FS}_{256}$} & \multicolumn{1}{c|}{$\mathcal{FS}_{260}$} & \multicolumn{1}{c|}{${\mathcal{US}}_{260}$} & \multicolumn{1}{c|}{$\mathcal{FS}_{256}$}& \multicolumn{1}{c|}{$\mathcal{FS}_{260}$} & \multicolumn{1}{c|}{${\mathcal{US}}_{260}$} \\ \hline \hline
			{MocoNet\cite{paper:moco}}  &   0.060 & 0.059 & 0.055 & 0.133 & 0.139& 0.117  \\  
			{Single-Branch}   & 0.013 & 0.015 & 0.013 & 0.020 & 0.043& 0.016\\  
			{Two-Branch}   & 0.012 & 0.013  & 0.011 & 0.019 & 0.043& 0.013\\ \hline
		\end{tabular}
	\end{center}
\end{table}


\section{Discussion}
The proposed motion-correction algorithm has been shown to correct for rigid in-plane motion, given the timing of motions. The method works for cases with multiple motions, and with a high degree of accuracy. Breaking up the $k$-space data into a dominant pose part, and the remaining poses part, enables the use of a data-consistency mechanism. The DC layer is used to ensure that a significant and uncorrupted part of $k$-space is left unchanged in the reconstructed image. Enforcing a data-consistency constraint has been successfully implemented with DNNs for the task of sparse MR image reconstruction~\cite{paper:schlemper}, where a constant under-sampling mask was used for the reconstruction task. In comparison, we generate a different, motion-timing dependent, mask for each case. The use of the motion timing to partition the $k$-space data and to impose a consistent $k$-space presents a novel approach to MRI motion correction using DNNs. 

The proposed method can simultaneously correct motion artefacts and reconstruct under-sampled MRI scans, given the appropriate scan order. Furthermore, our experiments show that our method is also able to correct for non-rigid and out-of-plane motions, as it mostly relies on the DP.

In summary, we present a novel approach for patient motion correction. The method is efficient and inexpensive, and seems to be a promising direction for bringing patient motion correction to the clinical practice. The availability of such a method will reduce the number of failed studies and improve the clinical efficiency of MR scanners.
\clearpage
\bibliographystyle{plain}
\bibliography{shortbib}
\end{document}